\documentclass[twocolumn,showpacs,preprintnumbers]{revtex4}
\usepackage{graphicx}% Include figure files
\usepackage{dcolumn}% Align table columns on decimal point
\usepackage{bm}% bold math
\usepackage{amssymb}
\usepackage{amsmath}
\usepackage{epsfig}    
\def\beq{\begin{equation}}
\def\eeq{\end{equation}}
\def\bea{\begin{array}}
\def\eea{\end{array}}
\def\be{\begin{equation}}
\def\ee{\end{equation}}
\def\ba{\begin{eqnarray}}
\def\ea{\end{eqnarray}}

\def\to{\rightarrow}

\def\[{\left[}
\def\]{\right]}
\def\({\left(}
\def\){\right)}

\def\sm0{{\widetilde{m}_0}}

\def\U1em{{U(1)_{\rm em}}}
\def\to{\rightarrow}

\def\sq2{\sqrt{2}}

\def\ee{e^+e^-}

\def\End{\end{document}}
\newcommand{\gsim}{\mbox{ \raisebox{-1.0ex}{$\stackrel{\textstyle >}
{\textstyle \sim}$ }}}
\newcommand{\lsim}{\mbox{ \raisebox{-1.0ex}{$\stackrel{\textstyle <}
{\textstyle \sim}$ }}}

\def\fsl#1{\setbox0=\hbox{$#1$}                 % set a box for #1 
   \dimen0=\wd0                                 % and get its size
   \setbox1=\hbox{/} \dimen1=\wd1               % get size of /
   \ifdim\dimen0>\dimen1                        % #1 is bigger
      \rlap{\hbox to \dimen0{\hfil/\hfil}}      % so center / in box
      #1                                        % and print #1
   \else                                        % / is bigger
      \rlap{\hbox to \dimen1{\hfil$#1$\hfil}}   % so center #1
      /                                         % and print /
   \fi}

\begin{document}                                                              
%\draft
%@twocolumnfalse\endcsname

\title{Neutrino mass, Dark Matter and Baryon Asymmetry\\
       via TeV-Scale Physics without Fine-Tuning}%
\author{%
{\sc Mayumi Aoki\,$^1$, Shinya Kanemura\,$^2$,
   and Osamu Seto\,$^3$}
}
\affiliation{%
%\address{\vspace*{5mm}
\vspace*{2mm} 
$^1$ICRR, University of Tokyo, Kashiwa 277-8582, Japan\\
$^2$Department of Physics, University of Toyama, 3190 Gofuku, Toyama 930-8555, Japan\\ 
$^3$Instituto~de~F\'{i}sica~Te\'{o}rica~UAM/CSIC,~Universidad~Aut\'{o}noma~de~Madrid, Cantoblanco, Madrid 28049, Spain\\
}
%\maketitle

%\vspace*{5mm} 
\begin{abstract}
%\hspace*{-0.35cm}
 We propose an extended version of the standard model, in which 
 neutrino oscillation, dark matter, and baryon asymmetry of 
the Universe can be simultaneously explained by the TeV-scale physics
 without assuming large hierarchy among the mass scales.
 Tiny neutrino masses are generated at the three loop level due to
 the exact $Z_2$ symmetry, by which stability of the dark
 matter candidate is guaranteed. The extra Higgs doublet is
 required not only for the tiny neutrino masses but also for
 successful electroweak baryogenesis.
 The model provides discriminative predictions especially in Higgs
 phenomenology, so that it is testable at current and future
 collider experiments.
\pacs{\, 14.60.Pq, 14.60.St, 14.80.Cp, 12.60.Fr  \hfill~~[\today] }
\end{abstract}
%\vskip 1pc]

\maketitle

\setcounter{footnote}{0}
\renewcommand{\thefootnote}{\arabic{footnote}}

%\section{Introduction} 

\begin{figure}[t]
\begin{center}
  \epsfig{file=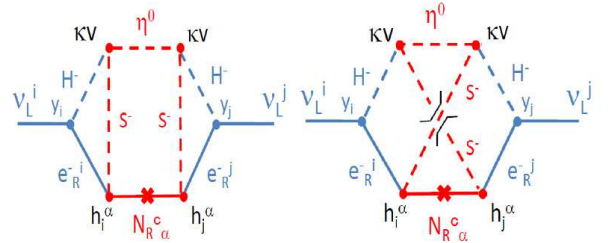,width=8cm}
\end{center}
  \caption{The diagrams for generating tiny neutrino masses. }
  \label{diag-numass}
\end{figure}

\begin{figure}[t]
\begin{center}
\epsfig{file=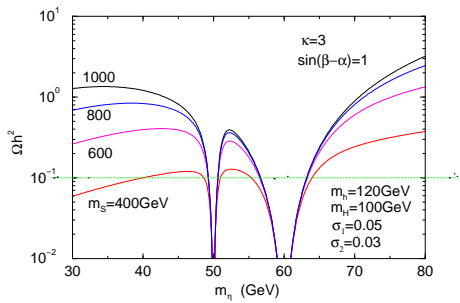,width=6.cm}
\end{center}
  \caption{The relic abundance of $\eta$. 
 }
  \label{etaOmega}
\end{figure}

\begin{figure}[t]
\begin{center}
  \epsfig{file=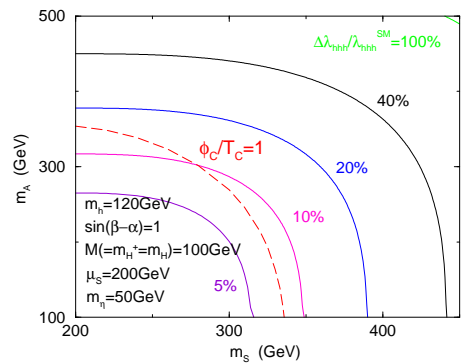,width=6.cm}
\end{center}
 \caption{The region of strong first order EWPT.
 Deviations from the SM value in the $hhh$ coupling
 are also shown. }
 \label{1opt}
\end{figure}

Although the standard model (SM) for elementary particles has been
successful for over three decades, the Higgs sector remains unknown.
The discovery of a Higgs boson is the most important issue
at the CERN Large Hadron Collider (LHC). 
On the other hand, today we have definite reasons to consider a model
beyond the SM.
First of all, the data indicate that neutrinos have tiny masses and mix
with each other\cite{lep-data}. 
Second, cosmological observations have revealed that the energy density of  
dark matter (DM) in the Universe dominates 
that of baryonic matter\cite{wimp}.
Finally, asymmetry of matter and anti-matter in our Universe has been addressed
as a serious problem regarding existence of ourselves\cite{sakharov}.
They are all beyond the scope of the SM, so that an extension of the SM is required
to explain these phenomena, which would be related to the physics of
electroweak symmetry breaking. 
\\ 
\indent
A simple scenario to generate tiny masses  ($m_\nu$) for left-handed
(LH) neutrinos would be
based on the seesaw mechanism with heavy right-handed (RH) neutrinos\cite{see-saw}; 
$m_\nu \simeq m_D^2/M_R$, where $M_R$ ($\sim 10^{13-16}$ GeV) is the
Majorana mass of RH neutrinos, and 
$m_D^{}$ is the Dirac mass of %at most
the electroweak scale.  
This scenario would be compatible with the framework with large mass
scales like grand unification.
However, introduction of such large scales causes a problem of hierarchy.
In addition, the decoupling theorem\cite{dec-theorem} makes it far from experimental tests.
\\
\indent
In this letter, we propose an alternative model which would explain neutrino
oscillation, origin of DM and baryon asymmetry simultaneously 
by an extended Higgs sector with RH neutrinos. 
In order to avoid large hierarchy, masses of the RH neutrinos are to be at most TeV scales. 
Tiny neutrino masses are then generated at the three loop level due to an
exact discrete symmetry, by which tree-level Yukawa couplings of neutrinos are prohibited.
The lightest neutral odd state under the discrete symmetry is a candidate of DM. 
Baryon asymmetry can be generated at the electroweak phase transition
(EWPT) by additional CP violating phases in the Higgs sector\cite{ewbg-thdm,ewbg-thdm2}.
In this framework, a successful model can be built without contradiction of the current data. \\
\indent
Original idea of generating tiny neutrino masses via the radiative effect 
has been proposed by Zee\cite{zee}. 
The extension with a TeV-scale  RH neutrino has been discussed in Ref.~\cite{knt},
where the neutrino masses are generated at the three-loop due to the exact $Z_2$
parity, and the $Z_2$-odd RH neutrino is a candidate of DM. This 
has been extended with two RH neutrinos
to describe the neutrino data\cite{kingman-seto}. 
%for the description of the neutrino data\cite{kingman-seto}. 
Several models with adding baryogenesis have been considered in
Ref.~\cite{ma}.
As compared to these models, the following new advantages are
in the present model:
(a)~all mass scales are at most at the TeV scale without large hierarchy, 
(b)~physics for generating neutrino masses is connected with that for
DM and baryogenesis, 
(c)~the model parameters are strongly constrained by the current data, so
    that the model gives discriminative predictions which can be tested
    at future experiments.
\\
 \indent
In addition to the {\it known} SM fields, particle entries are 
two scalar isospin doublets with hypercharge $1/2$ ($\Phi_1$ and $\Phi_2$),  
charged singlets ($S^\pm$), a real scalar singlet ($\eta$) and two
generation isospin-singlet RH neutrinos ($N_R^\alpha$ with $\alpha=1, 2$).
%At least two generations are necessary for $N_R^\alpha$ to reproduce the neutrino data.
%We here concentrate on the minimum case with $\alpha=1, 2$. 
%
In order to generate tiny neutrino masses
at the three-loop level,  
we impose an exact $Z_2$
symmetry as in Ref.~\cite{knt}, which we refer as $Z_2$. 
We assign the $Z_2$ odd charge to $S^\pm$, $\eta$ and $N_R^\alpha$, while 
ordinary gauge fields, quarks and leptons and Higgs doublets are  $Z_2$ even.
Introduction of two Higgs doublets would cause a dangerous flavor
changing neutral current. To avoid this in a natural way, we impose
another discrete symmetry  ($\tilde{Z}_2$) that is softly broken\cite{glashow-weinberg}.
From a phenomenological reason discussed later, we assign $\tilde{Z}_2$ charges such that
only $\Phi_1$ couples to leptons whereas $\Phi_2$ does to quarks;  
\begin{eqnarray}
 {\cal L}_Y  \!=\! -\!  y_{e_i}^{}  \overline{L}^i \Phi_1 e_R^i
     \! - \! y_{u_i}^{}  \overline{Q}^i \tilde{\Phi}_2 u_R^i
     \! - \! y_{d_i}^{}  \overline{Q}^i \Phi_2 d_R^i + {\rm h.c.}, \label{typex-yukawa}
\end{eqnarray}
where $Q^i$ ($L^i$) is the ordinary $i$-th generation LH quark (lepton)
doublet,  and $u_R^i$ and $d_R^i$ ($e_R^i$) are RH-singlet up- and
down-type quarks (charged leptons), respectively.   
We summarize the particle properties under 
$Z_2$ and $\tilde{Z}_2$ in TABLE~\ref{discrete}.
\begin{table}[t]
\begin{center}
  \begin{tabular}{c|ccccc|cc|ccc}
   \hline
   & $Q^i$ & $u_R^{i}$ & $d_R^{i}$ & $L^i$ & $e_R^i$ & $\Phi_1$ & $\Phi_2$ & $S^\pm$ &
    $\eta$ & $N_{R}^{\alpha}$ \\\hline
$Z_2\frac{}{}$                ({\rm exact}) & $+$ & $+$ & $+$ & $+$ & $+$ & $+$ & $+$ & $-$ & $-$ & $-$ \\ \hline  
$\tilde{Z}_2\frac{}{}$ ({\rm softly\hspace{1mm}broken})& $+$ & $-$ & $-$ & $+$ &
                       $+$ & $+$ & $-$ & $+$ & $-$ & $+$ \\\hline
   \end{tabular}
\end{center}
  \caption{Particle properties under the discrete symmetries.
 }
  \label{discrete}
 \end{table}
Notice that the Yukawa coupling in Eq.~(\ref{typex-yukawa})
is different from that in the minimal supersymmetric SM\cite{hhg}.
The scalar potential is given by
\begin{eqnarray}
&&\hspace*{-0.6cm} V = \sum_{a=1}^2 \left(-\mu_a^2 |\Phi_a|^2+\lambda_a|\Phi_a|^4 \right)
- (\mu_{12}^2 \Phi_1^\dagger \Phi_2 + {\rm h.c.}) \nonumber \\
&&\hspace*{-0.6cm}+\lambda_3|\Phi_1|^2|\Phi_2|^2 +\lambda_4
 |\Phi_1^\dagger \Phi_2|^2 + \left\{ \frac{\lambda_5}{2} (\Phi_1^\dagger
    \Phi_2)^2+ {\rm h.c.} \right\} \nonumber\\
 &&\hspace*{-0.6cm}    + \!\! \sum_{a=1}^2\!
  \left(\rho_a |\Phi_a|^2|S|^2 + \sigma_a |\Phi_a|^2
  \frac{\eta^2}{2} \right) \! +
  \!\!\! \sum_{a,b=1}^2\!\left\{ \kappa \,\,\epsilon_{ab} (\Phi^c_a)^\dagger
                    \Phi_b S^- \eta \right. \nonumber\\
 &&\hspace*{-0.6cm}
 \left. + {\rm h.c.}\right\}
+\!\mu_s^2 |S|^2+ \lambda_s |S|^4  +\! \mu_\eta^2 \eta^2/2 +
  \lambda_\eta \eta^4+ \!\xi |S|^2 \eta^2/2, % + V_{Z_2 {\rmodd}}(S^\pm, \eta),
 \end{eqnarray}
 where $\epsilon_{ab}$ is the anti-symmetric tensor with $\epsilon_{12}=1$.
The mass term and the interaction for $N_R^\alpha$ are given by 
\begin{eqnarray}
 {\cal L}_Y \!= \! \sum_{\alpha=1}^2\!\left\{ \!\frac{1}{2}m_{N_R^\alpha}^{} \overline{{N_R^\alpha}^c} N_R^\alpha
                 -  h_i^\alpha \overline{(e_R^i)^c}
                   N_R^\alpha S^-\! + {\rm h.c.}\!\right\}.
\end{eqnarray} 
\indent
In general, $\mu_{12}^2$, $\lambda_5$ and $\kappa$
(as well as $h_i^\alpha$) can be complex.
The phases of $\lambda_5$ and $\kappa$ can be eliminated by rephasing 
$S^\pm$ and $\Phi_1$. The remaining phase of $\mu_{12}^2$  
causes CP violation in the Higgs sector.
Although the phase is crucial for successful baryogenesis at the EWPT\cite{ewbg-thdm},
it does not much affect the following discussions. Thus, we neglect it for simplicity.
We later give a comment on the case with the non-zero CP-violating phase. \\
\indent
As $Z_2$ is exact, the even and odd fields cannot mix.
Mass matrices for the $Z_2$ even scalars are diagonalized as in the
usual two Higgs doublet model (THDM)
by the mixing angles $\alpha$ and $\beta$, where $\alpha$
diagonalizes the CP-even states, and
$\tan\beta=\langle \Phi_2^0
\rangle/\langle \Phi_1^0 \rangle$\cite{hhg}. 
The $Z_2$ even physical states are two CP-even ($h$ and $H$),
a CP-odd ($A$) and charged ($H^\pm$) states.
We here define $h$ and $H$ such that $h$ is always
the SM-like Higgs boson when $\sin(\beta-\alpha)=1$. \\
\indent
The LH neutrino mass matrix $M_{ij}$ is generated by the three-loop diagrams  in
FIG.~\ref{diag-numass}.
The absence of lower order loop contributions is guaranteed 
by $Z_2$.  $H^\pm$  and  $e_R^i$ play a crucial role to connect
LH neutrinos with the one-loop sub-diagram
by the $Z_2$-odd states.
We obtain
\begin{eqnarray}
M_{ij} = \sum_{\alpha=1}^{2} 
  C_{ij}^\alpha F(m_{H^{\pm}}^{},m_{S^{\pm}}^{},m_{N_R^{\alpha}}^{}, m_\eta), 
\end{eqnarray}
where $C_{ij}^\alpha =
   4 \kappa^2 \tan^2\!\beta 
  (y_{e_i}^{\rm SM} h_i^\alpha) (y_{e_j}^{\rm SM} h_j^\alpha)$ and
\begin{eqnarray}
&&\!\!\!\! F(m_{H^{\pm}}^{},m_{S^{\pm}}^{},m_{N_R}^{}, m_\eta) =
    \left(\frac{1}{16\pi^2}\right)^3 \frac{(- m_{N_R}^{}
    v^2)}{m_{N_R}^2-m_\eta^2} \nonumber\\
&&\!\!\!\! \times \int_0^{\infty} \! dx \left[  x   
   \left\{  \frac{B_1(-x,m_{H^{\pm}}^{},m_{S^\pm}^{})\!-\!
    B_1(-x,0,m_{S^{\pm}}^{})}{m_{H^\pm}^2} \right\}^2\right.\nonumber\\
&&\!\!\!\! \left.\times
 \left(\frac{m_{N_R}^2}{x+m_{N_R}^2}-\frac{m_\eta^2}{x+m_\eta^2}\right)
           \right],
\hspace{4mm} (m_{S^\pm}^{2}\gg m_{e_i}^{2}), 
\end{eqnarray}
with 
$m_f$ representing the mass of the field $f$, $y_{e_i}^{\rm
SM}=\sqrt{2}m_{e_i}/v$, $v\simeq 246$GeV and  
$B_{1}$ being the tensor coefficient function in Ref.~\cite{passarino-veltman}.
Magnitudes of 
$\kappa \tan\beta$ as well as $F$ determine the universal scale of $M_{ij}$, 
whereas variation of $h_i^\alpha$ ($i=e$, $\mu$, $\tau$) 
reproduces the mixing pattern indicated by the neutrino
 data\cite{lep-data}.
$M_{ij}$ is related to the data by
$M_{ij} = U_{is} (M_\nu^{\rm diag})_{st} (U^T)_{tj}$,
where $U_{is}$ is the unitary matrix and
$M_\nu^{\rm diag}$ $=$ ${\rm diag}(m_1, m_2, m_3)$.
\begin{table}
\begin{center}
  \begin{tabular}{|c||c|c|c|c|c|c|c|}\hline
     Set   & $h_e^1$ & $h_e^2$ & $h_\mu^1$ & $h_\mu^2$ & $h_\tau^1$ & $h_\tau^2$  &
   $B(\mu\!\!\to\!\! e\gamma)$\\\hline 
  A &  2.0    &  2.0     &  -0.019     & 0.042
                   &-0.0025   & 0.0012  &$6.9\!\times \!10^{-12}$\!\!\\\hline 
   B & 2.2     &  2.2     &  0.0085     & 0.038 
                   & -0.0012  &    0.0021     &$6.1\!\times \!10^{-12}$ \\\hline 
   \end{tabular}
\end{center}
 \caption{Values of $h_i^\alpha$ for $m_{H^\pm}^{} (m_{S^\pm}^{})=100(400)$GeV 
  $m_\eta=50$ GeV, $m_{N_R^1}=m_{N_R^2}=$3.0 TeV for the normal hierarchy. For  Set A (B), 
  $\kappa\tan\beta=28 (32)$ and $U_{e3}=0 (0.18)$. 
 Predictions on the branching ratio of $\mu\to e
 \gamma$ are also shown.}
  \label{h-numass}
 \end{table}
Under the {\it natural} requirement 
$h_e^\alpha \sim {\cal O}(1)$, and taking 
the  $\mu\to e\gamma$ search results into account\cite{lfv-data},   
we find that $m_{N_R^\alpha}^{} \sim {\cal O}(1)$ TeV, 
$m_{H^\pm}^{} \lsim {\cal O}(100)$ GeV, $\kappa \tan\beta \gsim {\cal
O}(10)$, and $m_{S^\pm}^{}$ being several
times 100 GeV. 
On the other hand, the LEP direct search results indicate 
$m_{H^\pm}^{}$ (and $m_{S^\pm}^{}$)  $\gsim 100$ GeV\cite{lep-data}.  
In addition, with the LEP precision measurement for the $\rho$ parameter,  
possible values uniquely turn out to be  
$m_{H^\pm}^{} \simeq m_{H}^{}$ (or $m_{A}^{}$) $\simeq 100$ GeV
for $\sin(\beta-\alpha) \simeq 1$. 
Thanks to the Yukawa coupling in Eq.~(\ref{typex-yukawa}), such
a light $H^\pm$ is not excluded by the $b \to s \gamma$ data\cite{bsgamma}.
Since we cannot avoid to include the hierarchy among $y_i^{\rm SM}$,  
we only require $h_i^\alpha y_i \sim {\cal O}(y_e) \sim 10^{-5}$ 
for values of $h_i^\alpha$. 
Several sets for $h_i^\alpha$ are shown in TABLE~\ref{h-numass} with the
predictions on the branching ratio of $\mu\to e\gamma$ 
assuming the normal hierarchy, %;
%{\it i.e.},
$m_1 \simeq m_2 \ll m_3$ with
$m_1=0$. 
For the inverted hierarchy ($m_3 \ll m_1\simeq m_2$ with $m_3=0$),   
$\kappa \tan\beta$ is required to be larger. 
Our model turns out to prefer the normal hierarchy
scenario\cite{comment1}. \\
%
%\section{Cold Dark Matter}
\indent
The lightest $Z_2$-odd particle is 
stable and can be a candidate of DM if it is neutral.
In our model, $N_R^\alpha$ must be heavy, so that 
the DM candidate is identified as $\eta$.
When $\eta$ is lighter than the W boson, $\eta$ dominantly annihilates 
into $b \bar{b}$ and $\tau^+\tau^-$ via tree-level $s$-channel
Higgs ($h$ and $H$) exchange diagrams, and into $\gamma\gamma$ via
one-loop diagrams.
From their summed thermal averaged annihilation rate $\langle \sigma v \rangle$,
the relic mass density  $\Omega_\eta h^2$ is 
evaluated as   
\begin{eqnarray}
 \Omega_\eta h^2= 1.1\times 10^9
%  \left.
\frac{(m_{\eta}/T_d)}{ \sqrt{g_\ast} M_P \langle \sigma
   v\rangle } 
%\right|_{T_d}
 \hspace{2mm} {\rm GeV}^{-1}, 
 \end{eqnarray}
where $M_P$ is the Planck scale, $g_\ast$ is the total number of
relativistic degrees of freedom in the thermal bath, and $T_d$ is the decoupling temperature\cite{kt}.
FIG.~\ref{etaOmega} shows 
$\Omega_{\eta}h^2$ as a function of $m_\eta$. 
Strong annihilation can be seen near $50$ GeV $\simeq m_H^{}/2$
($60$ GeV $\simeq m_h/2$) due to the resonance of $H$ ($h$) mediation.
The data ($\Omega_{\rm DM} h^2 \simeq 0.11$\cite{wimp}) indicate that $m_\eta$ is around 40-65 GeV. \\
%
%\section{Beryon Asymmetry}
\indent The model satisfies the necessary 
conditions for baryogenesis\cite{sakharov}.
Especially, departure from thermal equilibrium can be 
realized by the strong first order EWPT.
The free energy is given at a high temperature $T$ by\cite{hte} 
\begin{eqnarray}
 V_{eff}[\varphi, T]= D (T^2-T_0^2) \varphi^2 
                     - E T \varphi^3 
                     + \frac{\lambda_T}{4} \varphi^4 + ..., 
\end{eqnarray}
where $\varphi$ is the order parameter, and 
\begin{eqnarray}
E &\simeq& \frac{1}{12 \pi v^3} (6 m_W^3 + 3 m_Z^3  
 + m_A^3 + 2  m_{S^\pm}^3),\label{coefe}
\end{eqnarray}
with $D \simeq (6 m_W^2 + 3 m_Z^2 + 6 m_t^2 + m_{A}^2 + 2m_{S^\pm}^2 )/(24v^2)$, 
 $T_0^2 \sim m_h^2/(4D)$ and $\lambda_T\sim m_h^2/(2v^2)$. 
A large value of the coefficient $E$ 
is crucial for the strong first order EWPT\cite{ewbg-thdm2}.
In Eq.~(\ref{coefe}), quantum effects by $h$, $H$ and $H^\pm$ are neglected
since they are unimportant for $\sin(\beta-\alpha)\simeq 1$ and 
$m_{H^\pm}\simeq m_H^{}\simeq M \hspace{1mm}(\equiv
\sqrt{2\mu_{12}^2/\sin 2\beta})$ [the soft $\tilde{Z}_2$ breaking scale\cite{nondec}]. 
For sufficient sphaleron decoupling in the broken phase, it is required that\cite{sph-cond} 
\begin{eqnarray}
 \frac{\varphi_c}{T_{c}}  \left(\simeq \frac{2 E}{\lambda_{T_c}}\right) 
   \gsim 1, \label{sph2}
\end{eqnarray}
where $\varphi_c$ ($\neq 0$) and $T_c$ are the critical values of
$\varphi$ and $T$ at the EWPT.
In FIG.~\ref{1opt}, the allowed region under the condition of
Eq.~(\ref{sph2}) is shown. The condition is satisfied
when
$m_{S^{\pm}}^{} \gsim 350$ GeV 
for $m_A^{} \gsim 100$ GeV, 
$m_h \simeq 120$ GeV, $m_H^{} \simeq m_{H^\pm}^{} (\simeq 
M) \simeq 100$ GeV, $\mu_S^{}\simeq 200$ GeV and
$\sin(\beta-\alpha)\simeq 1$. Unitarity bounds are also satisfied
unless $m_A^{}$ ($m_S^{}$) is too larger than $M$ ($\mu_S^{}$)\cite{ewbg-thdm2,aks-full}. \\ 
\indent
A successful scenario which can simultaneously solve the above three issues 
under the data\cite{lep-data,lfv-data,bsgamma} would be 
\begin{eqnarray}
 \begin{array}{ll}
 \sin(\beta-\alpha) \simeq 1, &\!\! (\kappa \tan\beta) \simeq 30, \\
 m_h = 120 {\rm GeV},     &\!\! m_H^{} \simeq m_{H^\pm} (\simeq M) \simeq 100 {\rm GeV},    \\
 m_A \gsim 100 {\rm GeV},     &\!\! m_{S^\pm}^{}\sim 400{\rm GeV},\\
 m_{\eta} \simeq 40-65 {\rm GeV}, &\!\! m_{N_R^{1}} \simeq m_{N_R^{2}}
  \simeq 3 {\rm TeV}.\\
  \end{array} \label{scenario}
\end{eqnarray}
This is realized without assuming unnatural hierarchy among the
couplings. All the masses are between
${\cal O}(100)$ GeV and ${\cal O}(1)$ TeV. As they are required by the data, 
the model has a predictive power. We note that the masses of $A$ and $H$
can be exchanged with each other. \\
\indent
We outline phenomenological predictions in the scenario in~(\ref{scenario}) in
order.  The detailed analysis is shown elsewhere\cite{aks-full}.
(I)~$h$ is the SM-like Higgs boson, but decays into $\eta\eta$ when $m_\eta < m_h/2$.
        The branching ratio is about 36\% (25\%) for   $m_\eta \simeq 45$ (55) GeV.
        This is related to the DM abundance, so that  
         our DM scenario is testable at the LHC.
(II)~$\eta$  is potentially detectable
        by direct DM searches\cite{xmass},    
         because $\eta$ can scatter with nuclei 
        via the scalar exchange\cite{john}. 
(III)~For successful baryogenesis, the $hhh$ coupling
      has to deviate from the SM value by more 
        than 10-20 \%\cite{ewbg-thdm2} (see FIG.~\ref{1opt}), 
        which can be tested at the International Linear Collider (ILC)\cite{hhh-measurement}.
(IV)~$H$ (or $A$) can predominantly decay into $\tau^+\tau^-$ instead of $b\bar b$ for $\tan\beta\gsim 3$. 
        When $A$ (or $H$) is relatively heavy it can decay into $H^\pm W^\mp$ and $H Z$ (or $A Z$). 
(V)~the scenario with light $H^\pm$ and $H$ (or $A$) can be directly tested at the LHC
    via $pp\to W^\ast \to H H^\pm$ and $A H^\pm$\cite{wah}. 
(VI)~$S^\pm$ can  be produced in pair 
        at the LHC (the ILC)\cite{zee-ph}, and decay into
        $\tau^\pm \nu \eta$. 
        The signal would be a hard hadron pair\cite{hagiwara} with a
        large missing energy.
(VII)~The couplings $h_i^\alpha$ cause lepton flavor violation 
     such as $\mu\to e\gamma$ which would provide information
     on $m_{N_R^{\alpha}}$ at future experiments. \\
\indent
Finally, we comment on the case with the CP violating phases.
Our model includes the THDM, so that the same discussion can be applied in evaluation of  baryon number at the
EWPT\cite{ewbg-thdm}.  
The mass spectrum  would be changed to some extent, but most of the features
discussed above should be conserved with a little modification. 
%In addition, a possibility of a scenario with leptogenesis may also be
%considered.
\\
%
%\section{Discussions and Conclusions}
\indent
We have discussed the model solving neutrino oscillation, DM and baryon
asymmetry by the TeV scale physics without fine tuning. It gives
specific predictions in Higgs phenomenology, DM physics and flavor physics,
so that it is testable at current and future experiments. \\ 
%
%{\bf Acknowledgments}~~~\\[2mm]
\indent
This work was supported, in part, by Japan Society for the Promotion
of Science, Nos.~1945210 and  % (MA),
18034004, %(SK),
and the MEC project FPA 2004-02015, 
the Comunidad de Madrid project HEPHACOS (P-ESP-00346),  
and the European Network of Theoretical Astroparticle Physics
ILIAS/ENTApP (RII3-CT-2004-506222). %  (OS). 

%\vspace*{-4mm}


\begin{thebibliography}{1}
\bibitem{comment1}
The unitarity bound for the scattering process $h \eta \to H^+S^-$ turns
         out to give $\kappa \lsim 25$, whereas $\tan\beta$ should not be
         too large for successful electroweak baryogenesis\cite{ewbg-thdm}.
\end{thebibliography}
\end{document}